\newcommand{\onetype}[1]{\ensuremath{\mathbf{#1}}\xspace}

\newcommand{\timestamps}{\onetype{T}}

\documentclass[sn-basic]{sn-jnl}

\jyear{2021}%
\usepackage{xspace}

\usepackage{enumitem}
\usepackage[labelformat=empty, position=top]{subcaption}
\usepackage[export]{adjustbox}
\usepackage{fancyvrb}

\theoremstyle{thmstyleone}%
\theoremstyle{thmstyletwo}%

\theoremstyle{thmstylethree}%

\begin{document}

\title[Advancements and Challenges in Object-Centric Process Mining]{Advancements and Challenges in Object-Centric Process Mining: A Systematic Literature Review}


\author*[1]{\fnm{Alessandro} \sur{Berti}}\email{a.berti@pads.rwth-aachen.de}

\author[2]{\fnm{Marco} \sur{Montali}}\email{marco.montali@unibz.it}

\author[1]{\fnm{Wil M.P.} \sur{van der Aalst}}\email{wvdaalst@pads.rwth-aachen.de}

\affil*[1]{\orgdiv{Process and Data Science Group}, \orgname{RWTH Aachen University}, \orgaddress{\street{Ahornstrasse 55}, \city{Aachen}, \postcode{52074}, \state{NRW}, \country{Germany}}}

\affil*[2]{\orgdiv{Faculty of Engineering}, \orgname{Free University of Bozen-Bolzano}, \orgaddress{\street{Piazza Domenicani 3}, \city{Bolzano}, \postcode{39100}, \state{South Tyrol}, \country{Italy}}}

\abstract{
Recent years have seen the emergence of object-centric process mining techniques. Born as a response to the limitations of traditional process mining in analyzing event data from prevalent information systems like CRM and ERP, these techniques aim to tackle the deficiency, convergence, and divergence issues seen in traditional event logs. Despite the promise, the adoption in real-world process mining analyses remains limited. This paper embarks on a comprehensive literature review of object-centric process mining, providing insights into the current status of the discipline and its historical trajectory.
}


\maketitle

\section{Introduction}
\label{sec:introduction}

Process mining, as outlined in \cite{DBLP:books/sp/Aalst16}, forms an integral part of data science, aiming to extract valuable insights about business process executions from event data embedded in the supporting information systems. Traditional process mining methods, encompassing process discovery, conformance checking, model enhancement, and predictive analytics, hinge on a case notion — a criterion that bundles events belonging to a single execution of the business process.

Consider a ticketing management system; the ticket identifier consolidates all events related to its closure. Yet, certain information systems, like ERP, complicate the establishment of a single case notion. An invoice could link to multiple orders, and the same order may result in multiple invoices. Likewise, an order may be shipped through multiple packages, each of which may contain items from different orders. Such one-to-many and many-to-many relations make it unnatural, if not impossible, to view and analyze the process through a single, pre-defined case notion.

This is where object-centric process mining enters the scene. Tailored to analyze event data from such systems in a more intuitive manner, these techniques have seen foundational advancements in recent years. Only recently have they been implemented with publicly available tool support and event logs, owing to events not being tied to a single case notion, necessitating new data storage standards.

The discipline's infancy saw the proposal of substantially divergent methods for extraction, storage, preprocessing, discovery, conformance checking, and performance analysis. As such, navigating this wide range of approaches without expert guidance is a daunting task. Through this paper, we embark on a literature review journey to comprehend the primary research directions of this research line, provide a reading list of key results for beginners in the field, and offer insights into the open challenges and future steps. Furthermore, we aim to assess the discipline's maturity in terms of scalability and project support for analyzing event data in real-world information systems.

Guided by the principles laid out in \cite{DBLP:conf/ecis/BrockeSNRPC09}, our literature review follows five phases: defining the review scope, conceptualizing the topic, literature search and evaluation, literature analysis and synthesis, and setting the research agenda.

Following this introduction, the paper is structured as follows. Section \ref{sec:searchQueryAcceptanceCriteria} presents the search query and acceptance criteria (phases 1 and 3 of the methodology). Section \ref{sec:analysis} delves into an analysis of the results. This encompasses Subsections \ref{subsec:basicDefinitions}, \ref{subsec:models}, \ref{subsec:categoriesAnalysis}, \ref{subsec:paperDimensions}, and \ref{subsec:temporalEvolution} that respectively summarize the basic concepts and object-centric process models, propose a paper categorization, propose paper analysis dimensions, and describe the discipline's temporal evolution (phase 4). Finally, Section \ref{sec:discussion} concludes the paper with a reflection on the discipline's history and future outlook (phase 5).

\section{Search Query and Acceptance Criteria}
\label{sec:searchQueryAcceptanceCriteria}

In doing this literature review, we want to include any scientific paper on object-centric process mining.

\begin{table*}
\caption{Refinement process of the search query.}
\centering
\resizebox{0.9\textwidth}{!}{
\begin{tabular}{|p{9cm}|p{2cm}|}
\hline
{\bf Search Query} & {\bf Number of results} \\
\hline
TITLE-ABS-KEY("process mining") AND TITLE-ABS-KEY("object-centric" OR "artifact-centric") & 73 \\
\hline
TITLE-ABS-KEY("process mining"  OR "process discovery" OR "conformance checking") AND TITLE-ABS-KEY("object-centric" OR "artifact-centric") & 82 \\
\hline
TITLE-ABS-KEY("process mining"  OR "process discovery" OR "conformance checking") AND TITLE-ABS-KEY("object-centric" OR "artifact-centric" OR "object-aware") & 89 \\
\hline
TITLE-ABS-KEY("process mining"  OR "process discovery" OR "conformance checking") AND TITLE-ABS-KEY("object-centric" OR "artifact-centric" OR "object-aware" OR "multiple entities") & 94 \\
\hline
TITLE-ABS-KEY("process mining"  OR "process discovery" OR "conformance checking") AND TITLE-ABS-KEY("object-centric" OR "artifact-centric" OR "object-aware" OR "multiple entities" OR "graph database") & 114 \\
\hline
TITLE-ABS-KEY("process mining"  OR "process discovery" OR "conformance checking") AND TITLE-ABS-KEY("object-centric" OR "artifact-centric" OR "object-aware" OR "multiple entities" OR "graph database" OR "case notion") & 119 \\
\hline
TITLE-ABS-KEY("process mining"  OR "process discovery" OR "conformance checking") AND TITLE-ABS-KEY("object-centric" OR "artifact-centric" OR "object-aware" OR "multiple entities" OR "graph database" OR "case notion" OR "merging event logs") OR (TITLE-ABS-KEY("object-aware") AND TITLE-ABS-KEY("petri net")) & 127 \\
\hline
\end{tabular}
}
\label{tab:refinementSearchQuery}
\end{table*}

The refinement of the search query, illustrated in Table \ref{tab:refinementSearchQuery}, utilized the \emph{scopus} search engine and incorporated an iterative approach with various keywords related to process mining, object-centric, and artifact-centric concepts.

Initially, we explored various synonyms of artifact-centric and object-centric terms, including \emph{object-aware}, \emph{multi-case}, \emph{multi-instance}, \emph{multiple cases}, \emph{multiple instances}, and \emph{multiple process instances}. However, only \emph{object-aware} and \emph{multiple entities} significantly refined the search query, yielding additional pertinent results.

The potential of \emph{graph databases} for storing and querying object-centric event data is noteworthy, given their capability for handling extensive interconnections (event-to-event, event-to-object, object-to-object). Despite testing different synonyms such as event graph, knowledge graph, and property graph, these terms did not supplement the original query with additional findings, hence, they were omitted from the final search query.

Historically, before the advent of object-centric process mining techniques, the choice of a \emph{case notion} was a fundamental step in the analysis process. Consequently, the inclusion of techniques assisting in the selection of the case notion provides valuable historical context to this review.

Likewise, \emph{merging techniques} that amalgamate event data from various systems or organizations played a pivotal role before object-centric techniques were available. By including log merging techniques in this review, we aim to gain comprehensive insights into the handling of multiple event data sources in the past.

Finally, we considered it necessary to include \emph{object-aware extensions of the Petri net concept} in our search query to shed light on those works that infuse the control flow specification of processes (typically tackled with Petri nets) with objects, relations, and their manipulation. This, in turn, serves as a reference for forthcoming advancements in the field.

The final query is reported below:

\begin{Verbatim}[frame=single]
(TITLE-ABS-KEY("process mining"  OR "process discovery"
OR "conformance checking")
AND TITLE-ABS-KEY("object-centric" OR "artifact-centric"
OR "object-aware" OR "multiple entities"
OR "graph database" OR "case notion" OR "merging event logs"))
OR (TITLE-ABS-KEY("object-aware")
AND TITLE-ABS-KEY("petri net"))
\end{Verbatim}

The search query produced some spurious results, so we identified some acceptance criteria to filter the results in agreement with the goals of this review:
\begin{enumerate}
\item The reference should be properly inserted and mentioned in \emph{scopus}. Moreover, for the duplicated entries, only one entry is kept. This reduces the number of results from {\bf 127} to {\bf 96}.
An example of an entry excluded with this criteria is \emph{28th International Conference on Cooperative Information Systems, CoopIS 2022} because it (wrongly) reports the name of a conference as
a reference to a specific paper.
\item The event data or process model used in the paper should allow for the interaction of an event with several objects of different types, the interaction between cases of different event logs,
the choice of a case notion between many available possibilities, or being about advanced object-aware processes on top of which process discovery or conformance checking techniques
are currently in the work. This reduces the number of results from {\bf 96} to {\bf 71}.
Examples of entries excluded with this criteria are
\emph{Uncertain Case Identifiers in Process Mining: A User Study of the Event-Case Correlation Problem on Click Data}\footnote{Pegoraro, Marco, et al. "Uncertain case identifiers in process mining: A user study of the event-case correlation problem on click data." Enterprise, Business-Process and Information Systems Modeling: 23rd International Conference, BPMDS 2022 and 27th International Conference, EMMSAD 2022, Held at CAiSE 2022, Leuven, Belgium, June 6–7, 2022, Proceedings. Cham: Springer International Publishing, 2022.}
(because the event correlation is done on data without any case identifier) and \emph{Graph-Based Token Replay for Online Conformance Checking}\footnote{Waspada, Indra, et al. "Graph-Based Token Replay for Online Conformance Checking." IEEE Access 10 (2022): 102737-102752.}
(because a traditional token-based replay technique is applied).
\item The paper was published between January 2010 and June 2023 (this does not reduce the number of results, but makes the review reproducible).
\end{enumerate}

\section{Analysis}
\label{sec:analysis}

In this section, we provide a general framework for classifying the reviewed papers, on the one hand considering the modeling dimensions when capturing object-centric event data and corresponding processes, and on the other focusing on the different types of analysis. We start the required preliminaries on the modeling dimensions. We continue with a categorization of the papers obtained with the search query and with the core section that reviews the papers using our classification scheme. We complete with a brief account of the temporal evolution of the discipline.

\subsection{Fundamental Definitions}
\label{subsec:basicDefinitions}

The execution of business processes is supported by information systems, such as database management systems.
Such information systems contain relevant information about different \emph{objects}, also called entities. As customary in every data modeling approach, objects belong to different \emph{object types} (people, places, documents), also called classes.

Also, information systems store information about the \emph{events} related to the creation \allowbreak / \allowbreak deletion \allowbreak / \allowbreak update of an object or a set of objects, resulting from the execution of one or of multiple processes.
We could group events in \emph{event types}, also called \emph{activities}, based on their logical function (for example, we could group all the events creating an object of type person
into the class \emph{create person record}).

As customary in process mining, regardless of its specific type, every event happens at a specific point in time, called the \emph{timestamp} of the event. Each timestamp belongs to a dedicated type \timestamps, which in turn is an infinite set equipped with a total order. Concrete representations for \timestamps are, for example, integer or real numbers. 

The \emph{lifecycle} of an object is the set of events related to the object (for example, all the events going from the creation/birth to the deletion/death of the person).
Assuming a total order on the events of the information systems (e.g., that induced by event timestamps), each object with a non-empty lifecycle is associated with a \emph{start event} and an \emph{end event}.
The term \emph{artifact} is also used to refer to relevant objects and their lifecycle.

Objects and events may participate in different types of (binary) relationships, also stored inside the information system, and constituting an integral part of (object-centric) event logs. 
Three categories of relationship types are distinguished in the literature: 
\begin{itemize}
\item {\bf E2O} - relating events to possibly multiple objects. A pair $(e, o)$ conceptually indicates that event $e$ is related to object $o$. In general, events may be related to objects of distinct types. As an example, two {\bf E2O} relationships can be used to indicate that \emph{add item} events relate to a container order and to a target item (likewise for \emph{remove item}).  
\item {\bf O2O} - relating objects to other objects. A pair $(o_1, o_2)$ conceptually indicates that the object $o_1$ is related to $o_2$. An example is the containment relationship connecting an order to its items. 
\item {\bf E2E} - relating events to other events, e.g., to capture events with different levels of granularity or to express control flow relations (such as start-complete, directly-follows, \ldots) among events of the same granularity.
\end{itemize}
Additional key aspects, not discussed here, are needed to further detail the aforementioned modeling elements. In particular, relationship \emph{qualifiers} can be used to specify the meaning of relationships, or, in the {\bf E2O} case, to indicate how events operate over their connected objects. For example, one may want to capture that \emph{add item} has the effect of creating an {\bf O2O} containment relationship between the container order and the target item (while \emph{remove item} has the effect of ceasing the containment relationship).  Also, (timed) attributes can be used to specify properties of objects/events and how they vary over time. Furthermore, notice that additional modeling concerns (such as n-ary relationships and properties of relationships) can be seamlessly captured using the elements introduced before by \emph{reifying} relationship instances into corresponding dedicated objects.

\emph{Traditional event logs} record objects of a single object type
and the events are associated with exactly one object.
We call \emph{case} a set of events belonging to the same execution of the process. A \emph{case notion} is a criterion used to group events into a case.
\emph{Object-centric event logs} drop these assumptions, so objects of different object types are recorded. Different types of object-centric event logs have been proposed in the literature, and we could identify some interesting combinations.

For example, the popular OCEL format \url{https://www.ocel-standard.org} covers {\bf E2O} (where relationships are qualified at all), but not  {\bf O2O} nor {\bf E2E} relationships.
The commercial vendor Celonis recently proposed the usage of multi-event logs, which are collections of event logs and relationships between the cases/objects
of such logs. Therefore, multi-event logs belong to the category {\bf O2O}. Celonis also proposed the signal link functionality exploiting event-to-event relationships ({\bf E2E}) to connect event data \url{https://docs.celonis.com/en/network-explorer-and-signal-link-explorer.html}. Approaches based on knowledge graphs \cite{DBLP:conf/icpm/Xiong0KMGC22} and so-called \emph{event knowledge graphs} \cite{DBLP:books/sp/22/Fahland22} capture a mixture of E2O, O2O, and possibly E2E relationships in their full generality. 

\begin{figure}[ht]
\centering
\fbox{\includegraphics[width=\textwidth]{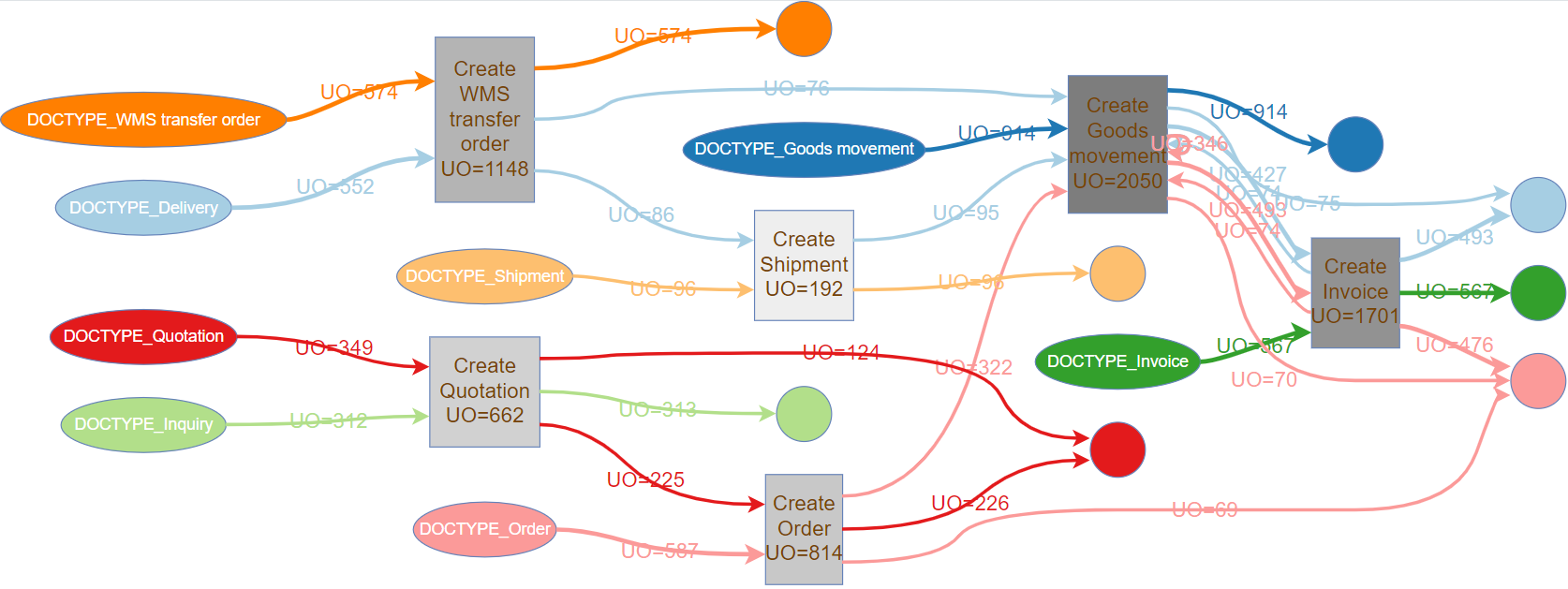}}
\caption{Object-centric directly-follows graph from \cite{DBLP:journals/sttt/BertiA23}. This process model belongs to the {\bf ET2ET} (activities are connected by arcs) and {\bf ET2OT} (arcs are colored according to the object type) categories.
}
\label{fig:ocdfgExample}
\end{figure}

The act of \emph{flattening} an object-centric event log into a traditional event log necessitates the selection of a case notion. This choice inherently leads to three primary issues: \emph{deficiency}, \emph{convergence}, and \emph{divergence} as articulated in \cite{DBLP:conf/sefm/Aalst19}.
The problem of \emph{deficiency} arises when certain events fail to be encompassed within any case. For instance, events linked to a purchase requisition that does not culminate into a purchase order may go unnoticed if the order is selected as the case notion.
Conversely, the issue of \emph{convergence} materializes when it becomes necessary to duplicate the same event across various cases. A classic illustration of this would be the events associated with an invoice comprising multiple orders when the order is elected as the case notion.
Lastly, the \emph{divergence} problem emerges when multiple instances of the same activity are presented within a single case. A case in point would be the events correlated with different invoices belonging to the same order.

\subsection{Object-Centric Process Models}
\label{subsec:models}

\begin{figure}[ht]
\centering
\fbox{\includegraphics[width=\textwidth]{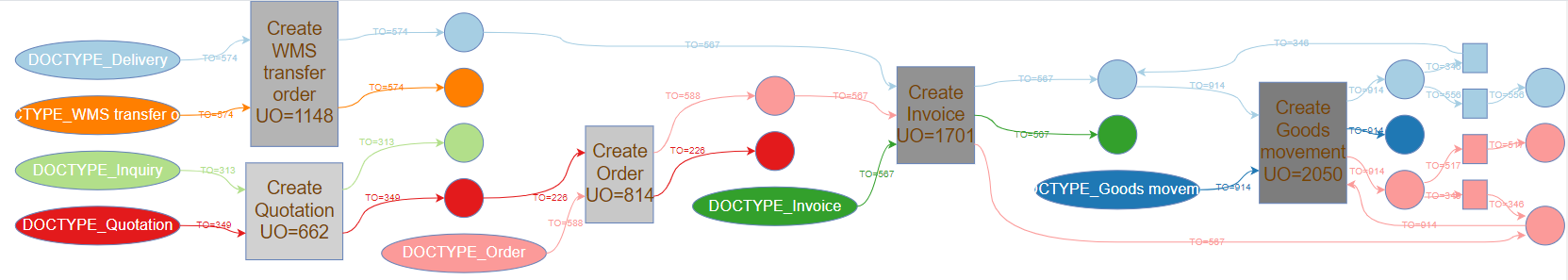}}
\caption{Object-centric Petri net from \cite{DBLP:journals/fuin/AalstB20}. This process model belongs to the {\bf ET2ET} (activities are indirectly connected by means of Petri net places) and {\bf ET2OT} (arcs are colored according to the object type) categories.}
\label{fig:ocpnExample}
\end{figure}

\begin{figure}[ht]
\centering
\fbox{\includegraphics[width=\textwidth]{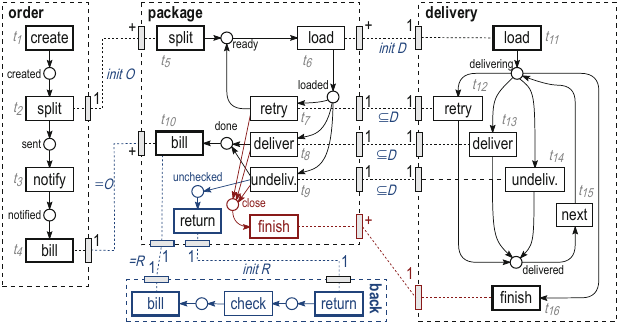}}
\caption{A Proclet from \cite{DBLP:conf/apn/Fahland19}. This process model shows features of all the three categories: {\bf ET2ET} (activities are indirectly connected through Petri net places), {\bf ET2OT} (each activity belongs to an object type and may synchronize with activities in other object types), and {\bf OT2OT} (synchronized activities implicitly establish relationships that can be used to constraint which objects can simultaneously flow through them).}
\label{fig:procletExample}
\end{figure}

\begin{figure}[ht]
\centering
\fbox{\includegraphics[width=\textwidth]{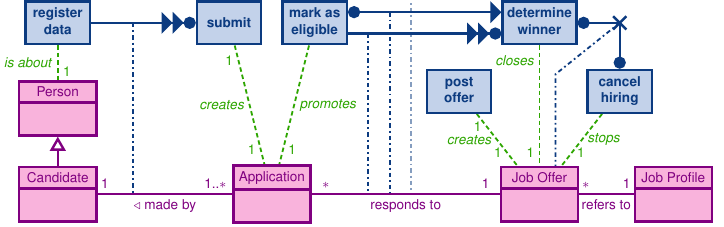}}
\caption{Object-centric behavioral constraints model from \cite{DBLP:conf/birthday/ArtaleCMA19}. process model shows features of all the three categories: {\bf ET2ET} (activities are connected by temporal constraints), {\bf ET2OT} (activities are linked to object types), and {\bf OT2OT} (object types are linked by UML associations).}
\label{fig:ocbcExample}
\end{figure}

\begin{figure}[ht]
\centering
\fbox{\includegraphics[width=\textwidth]{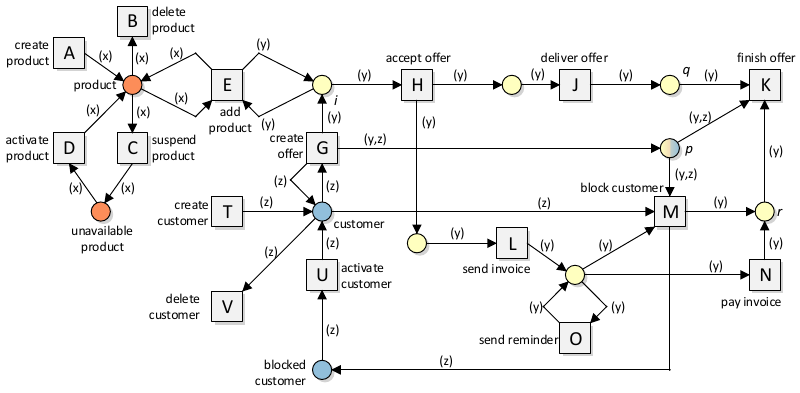}}
\caption{Petri net with identifier from \cite{DBLP:conf/apn/WerfRPM22}. process model shows features of all the three categories: {\bf ET2ET} (activities are indirectly connected through Petri net places), {\bf ET2OT} (activities manipulate objects carried by tokens), and {\bf OT2OT} (tokens can carry tuples of objects).}
\label{fig:pnidExample}
\end{figure}

Object-centric process mining techniques exploit object-centric event logs and \emph{object-centric process models}.
Object-centric process models can be abstracted as multigraphs in which the set of nodes are the activities or the object types involved in the business process, and
logical connections of different types connect the nodes. 
To classify object-centric process modeling languages, we mirror what is discussed for object-centric event logs, obtaining the following categories:
\begin{itemize}
\item {\bf ET2ET}: \emph{the model supports logical connections between event types/activities.} These connections can either be direct (e.g., two activities are connected in a directly-follows graph, see Fig. \ref{fig:ocdfgExample}) or indirect (e.g., there is a path in the Petri net between two visible transitions, see Fig. \ref{fig:ocpnExample}).
\item {\bf ET2OT}: \emph{the model supports for logical connections between event types/activities and object types/classes.}
These connections can be reported as arcs (between event and object types) or as arc types
(examples: activity-to-activity arcs, see Fig.~\ref{fig:ocdfgExample}, Fig.~\ref{fig:ocpnExample} and Fig.~\ref{fig:pnidExample}).
\item {\bf OT2OT}: \emph{the model allows for logical connections between object types/classes.} These connections are often explicit (e.g., shown using conventional representations from data modeling notations) and augmented with conditions on the start/end/synchronization between object types (see Fig.~\ref{fig:procletExample}) or cardinality constraints (see Fig.~\ref{fig:ocbcExample}).
\end{itemize}

A variety of modeling languages provide constructs within such three main categories. We briefly describe the ones that are being tackled by process mining techniques.

\emph{Object-centric directly-follows graphs} \cite{DBLP:journals/sttt/BertiA23} and object-centric Petri nets \cite{DBLP:journals/fuin/AalstB20} extend DFGs and Petri nets to the case where multiple objects flow through the process, and activities may either operate on a single or different objects of possibly different types. This is done by typing states/arcs with the type of objects that can reside in/flow through them, which is graphically rendered with different colors used for different types. Two examples are given in Fig.~\ref{fig:ocdfgExample} and Fig.~\ref{fig:ocpnExample}. 

These two approaches provide {\bf ET2ET} and {\bf ET2OT} constructs but do not allow for object-to-object relationships, which in turn makes it impossible to define conditions for execution requiring that the objects that simultaneously flow through the same activity must be in a certain relationship. This is essential in Petri nets to realize object-centric forms of synchronization, e.g., the fact that an order can only be shipped once \emph{all its contained items} have been validated. 

More advanced modeling languages have been proposed to incorporate {\bf OT2OT} constructs and, in turn, deal with such advanced forms of synchronization - with the downside that they are more difficult to use within process mining, and only a few process mining approaches have dealt with them up to now.

Proclet (in particular in the variant presented in \cite{DBLP:conf/apn/Fahland19}) and object-centric behavioral constraint (OCBC) models \cite{DBLP:conf/bis/LiCA17,DBLP:conf/bpm/ArtaleKMA19} do so through an implicit form of object manipulation, in the sense that objects/relationships are implicitly created by activities, depending on the constraints attached to them. 

Specifically, Proclets are built starting from object-specific Petri nets, each capturing the lifecycle of an object type in the process. Transitions from different lifecycles are synchronized by introducing one-to-one and one-to-many relationships, which indicate how many objects of each type can/must flow together through the transition. This implicitly establishes a relationship between such objects, which can be later used for different forms of synchronization. For example, in Fig.~\ref{fig:procletExample}, one can see that an order is split into multiple packages. When this happens, the so-called correlation set $O$ is initialized, recalling the order identifier and all the package identifiers created therein. Later, billing uses $O$ to indicate that an order can be billed only if all its packages are contingently billed as well.

Differently from Proclets, OCBC models employ declarative constraints that co-refer through object and relationship types in a data model. Such co-reference is used to scope the effect of each constraint. In the case of a co-reference on an object type, the constraint is applied to activities executed on the same object. In the case of co-reference on a relationship type, the constraint is applied to pairs of objects connected through that relationship type. This must be substantiated in temporal semantics that simultaneously deals with activities, objects, and their relationships \cite{DBLP:conf/bpm/ArtaleKMA19}. For example, in Fig.~\ref{fig:ocbcExample}, the constraint insisting on the \emph{submit} and \emph{register data} activities indicates that an application can only be submitted once the candidate owning that application has registered their own data.

A different approach to deal with all the three listed categories is by enhancing Petri nets with identifiers, on the one hand dictating that tokens can carry a single identifier (representing an object) or a tuple of identifiers (representing a relationship), and on the other providing inscriptions on arcs that are used to create, consume, and manipulate tokens and the identifiers they carry. For example, in Fig.~\ref{fig:pnidExample}, creating an offer requires a customer $z$ in the \emph{customer} place and has the effect of creating a new offer $y$, and of establishing a relationship $(y,z)$ between the newly created offer $y$ and the customer $z$. The new offer $y$ is inserted in place $i$, while the pair $(y,z)$ is stored in place $p$. This approach provides the backbone for different models of object-centric processes where additional constraints are imposed on such identifiers \cite{DBLP:conf/caise/PolyvyanyyWOB19,DBLP:journals/is/GhilardiGMR22}. Notably, \cite{DBLP:journals/is/GhilardiGMR22} highlights how this approach relates to Proclets and the different sychronization schemes introduced in \cite{DBLP:conf/apn/Fahland19}. 

As a final remark, we close by recalling that the term \emph{artifact-centric} is commonly used to refer to process models that explicitly deal with the {\bf ET2ET} and {\bf OT2OT} categories, and embed {\bf ET2OT} features inside the effects of activities. Notable examples in this spectrum are GSM \cite{DBLP:conf/debs/HullDMFGHHHLMNSV11} and BAUML \cite{DBLP:conf/cikm/CalvaneseMET14}.

\subsection{Categories of Analysis}
\label{subsec:categoriesAnalysis}

In this subsection, we categorize the results obtained from the search query into different categories.
Table \ref{tab:includedResults} includes a synthetic representation of such categories.

\subsubsection{Extraction of Object-Centric Event Data}
The techniques described in this subsection allow us to connect to a relational/non-relational database and extract an object-centric event log.

Here, we distinguish between the definition of meta-models for the extraction (without any proposed implementation, e.g., \cite{DBLP:journals/ijdsst/PajicB16,simovic2018domain,pajic2021towards})
or actual extraction techniques that have been used to extract an object-centric event log.
In \cite{DBLP:journals/tsc/LuNWF15}, an approach for artifact-centric process mining on top of SAP ERP is proposed, with an implementation (as a plug-in of ProM) allowing to ingest the artifact-centricity from SAP and visualize the performance
of the artifacts. According to the authors, the implementation is tricky to configure and this makes the usage difficult for the final user.
In \cite{DBLP:conf/icpm/BertiPRA21,DBLP:conf/icpm/WeberPRA22}, a generic approach to extract event data from SAP ERP is proposed. This starts from the specification of a starting object type/table of SAP and its progressive extension
to a bigger set of object types/tables through user interaction. Then, the tables and the relationships between them are considered to automatically extract an object-centric event log.
This is user-friendly (no SQL query is asked to the end user), allows to extract quickly different processes in SAP, and the tool support is publicly available. However, the paper does not properly assess the resulting object-centric event log.
In \cite{DBLP:conf/icpm/VugsAS22}, Konetki, a tool offering a structured approach to prepare event data for process mining, is presented. Konetki implements an object-centric data model from which object-centric event logs can be extracted
or different traditional event logs can be exported using different case notions.

The extraction of an event log from the Dollibar ERP system is described in \cite{DBLP:conf/caise/LiMCA18}. This log can be used with the object-centric behavioral constraints techniques.
Also, a log (always for OCBC techniques) has been extracted starting from popular social media platforms (Facebook, Linkedin, Twitter) in \cite{DBLP:journals/access/LiC19c}.

In \cite{DBLP:conf/dlog/Xiong0KMGC22,DBLP:conf/icpm/Xiong0KMGC22}, the Virtual Knowledge Graph (VKG) approach to access data in relational databases is exploited to extract OCEL logs from databases.
The VKG approach is knowledge-based (and a similar approach, OnProm, has been previously proposed to extract XES logs from databases).

\subsubsection{Storage of Object-Centric Event Data}

Here, we introduce the different formats that have been proposed for the storage of object-centric event data.

For object-centric event logs satisfying {\bf E2O}, a tabular format has been proposed in \cite{DBLP:conf/simpda/BertiA19} and \cite{DBLP:conf/sefm/Aalst19}. Each row corresponds to an event,
the attributes at the event level are columns, and also the object types are columns.
Each cell corresponding to an object type column contains the set of related objects to the event of such an object type.
Eventually, the OCEL specification has been proposed in \cite{DBLP:conf/adbis/GhahfarokhiPBA21} with some implementations in JSON and XML.
The standard allows for attributes at the event and object level; moreover, each event is related to a set of objects.

For object-centric event logs satisfying {\bf O2O}, the ACEL standard has been proposed in \cite{DBLP:conf/IEEEscc/Moctar-MBabaASG22,m2023process} for the storage of artifact-centric event data
from blockchain applications. On a note, the ACEL meta-model requires significant modifications of the OCEL meta-model.

The XOC format that has been proposed in \cite{DBLP:conf/caise/LiMCA18} to store object-centric event data satisfying {\bf E2O} and {\bf O2O}.
This can be exploited for the discovery of object-centric behavioral constraint models. Every event that changes the status of the database model (creation, update, deletion of rows in the tables) is stored along
with the entire status of the database (since the discovery of object-centric behavioral constraint models requires such information).
This results in large logs even for small databases (because of the persistence of the status of the database for every event).

Some alternative storages have been adopted in the context of object-centric event data. In particular, graph databases \cite{DBLP:conf/bpm/EsserF19,DBLP:journals/jodsn/EsserF21,DBLP:conf/coopis/EldinAKBG22} have been used to store and query object-centric event data.
Also, document databases are used to store object-centric event data (in particular, MongoDB \cite{DBLP:journals/corr/abs-2202-05639}). The Parquet columnar format has been proposed in \cite{DBLP:conf/simpda/BertiA19} for keeping object-centric event data
in-memory. As efficient compression algorithms can be used on the columns of a columnar database,
the storage proves enormously competitive in comparison to OCEL for basic object-centric event data but loses the possibility to store nested attributes.

{\bf Limitations of the Current Formats}:
The importance of storing dynamic object attribute values has been highlighted as a necessity in two recent studies. In the context of Internet of Things (IoT),
the study \cite{DBLP:conf/bpm/BertrandWS22} evidences the need for accommodating rapidly changing IoT data within process mining techniques, underscoring the inability of existing formats like OCEL to track dynamic attribute values effectively.
The study \cite{DBLP:conf/icpm/GoossensSVA22}, concentrating on data-aware process mining, points out the ambiguity problems in linking attributes to events or objects in current object-centric event log formats, emphasizing the importance of support for dynamic
object attributes.

\subsubsection{Preprocessing of Object-Centric Event Data}

We consider in this section only techniques that prepare/transform data for the application of other techniques (not the internal preprocessing steps of every algorithm).

In this section, we divide between merging techniques (some event logs are provided as input, and an event log is obtained as output containing the information of the different logs), the choice of a case notion
(a criterion that is used to group events together) and filtering techniques (allowing to get an object-centric event log with a subset of the behavior of the original log).

{\bf Merging event logs}: different approaches have been proposed to merge the contribution of different event logs into one. In \cite{DBLP:conf/bpm/RaichelsonS14},
an approach to merge event logs with many-to-many relationships is proposed.
A more user-guided technique is proposed in \cite{DBLP:journals/eswa/ClaesP14}.
The usage of a hybrid artificial immune algorithm for event log merging is proposed in \cite{xu2016merging}.
Since the merged logs may have different levels of granularity, in \cite{DBLP:journals/is/RaichelsonSV17} an approach to consider such granularity is proposed.
Therefore, merging allows the application of traditional process mining techniques to a set of interconnected logs. This is of particular importance for inter-organizational processes.
In \cite{DBLP:conf/smds/Aalst21}, federated process mining is proposed to create an event log out of event data collected across organizational boundaries.
Another approach for merging event logs in the context of inter-organizational process mining is proposed in \cite{hernandez2021merging}.

{\bf Case Notion:} we assume that events can be extracted from an information system with many different attributes and connections to objects, but it is unclear which are the suitable case notions in order
to use the event data with traditional process mining techniques. Implementation of the OpenSLEX meta-model \cite{DBLP:conf/bpm/MurillasAR15,DBLP:journals/kais/MurillasRA20}
act as a bridge between the database and the eventual event logs.
The contents of the database are transformed into a format that allows the user to easily query the data and retrieve a list of possible case notions.
This can then be used to construct a traditional event log.

{\bf Filtering:} the tool proposed in \cite{DBLP:journals/sttt/BertiA23} allows some basic filtering operations (activities occurrences, number of objects per object type, sampling, activity - object type filtering).
Also, the MongoDB support for OCEL \cite{DBLP:journals/corr/abs-2202-05639}
naturally allows for many filters on the event data. Graph databases usually have powerful query languages. In \cite{DBLP:conf/coopis/EldinAKBG22}, the Neo4J graph database language Cypher is used to preprocess the
event data before other operations, such as the discovery of a process model, are used.

{\bf Feature Extraction}: in \cite{DBLP:conf/icpm/AdamsSSSA22}, the interconnections between different objects in an object-centric event log are used to define variants in the object-centric setting.
In particular, the connected components of the object interaction graph are used to identify clusters of related objects. These clusters are used to identify
correlated events in the object-centric event log. The graph-related properties of these events and objects are then used to identify events/objects with similar behavior
(this is eventually called variant).
These properties are also exploited in \cite{DBLP:conf/icsoc/AdamsPLSA22} for feature extraction on object-centric event logs.
Different use cases (including predictive analytics) and types of feature extraction (table-based, graph-based) are proposed.
In particular, the importance of the single features for the target is identified using explainable AI techniques (including the usage of SHAP values).
The assessment of the paper is done starting from a traditional event log ``adapted'' as an object-centric event log. The related tool support is publicly available and can be easily installed.
The paper \cite{DBLP:journals/eswa/GalantiLNM23} introduces feature extraction on object-centric event logs for the goals of predictive analytics (prediction of the remaining time, activities occurrences, and customer satisfaction).
It shows how considering the interaction between objects (object interaction graph) leads to an improvement in the quality of the prediction on object-centric event logs,
compared to only considering the flattened event logs. This has been evaluated in a real-life case study.
Also, \cite{DBLP:conf/icsoc/GherissiHG22} considers predictive monitoring starting from object-centric event logs, focusing on the prediction of the next activity and remaining time
starting from traditional and interaction-based features.

{\bf Alternative Approaches}: in \cite{DBLP:conf/bpm/RebmannRA22}, an approach to transform traditional event logs into OCEL(s) is proposed.
In \cite{DBLP:journals/is/KumarST23}, an approach to normalize an object-centric event log to a STAR and fully normalized database schema is shown.
In \cite{faria2022clustering}, an object-centric clustering approach with the purpose of identifying alternative ``behavioral patterns'' is proposed, helping the user to understand different executions of the same process.
Clustering is also proposed in \cite{DBLP:journals/access/Jalali22}, but with the different purpose of identifying object types that would produce a similar flattened traditional event log.

\subsubsection{Process Discovery Techniques}

Here, we divide between techniques discovering models satisfying {\bf ET2ET+OT2OT} (but not {\bf ET2OT}), {\bf ET2ET+ET2OT} (but not {\bf OT2OT}) and {\bf ET2ET+OT2OT+ET2OT}.

{\bf ET2ET+OT2OT}: in \cite{DBLP:conf/bpm/NooijenDF12}, a set of discovery techniques is proposed starting from the relational database. First, the artifact schema is discovered from the relational database.
Then, the data is divided between artifacts. This data is used to create event logs for the single artifacts, and to discover the lifecycle of the artifacts.
An approach for artifact-centric process discovery specific to SAP ERP is proposed in \cite{DBLP:journals/tsc/LuNWF15}. Moreover, a meta-model that can be used to extract an event log
from generic ERP systems is described in \cite{DBLP:journals/ijdsst/PajicB16}, with an application on Dynamics NAV ERP.

A discovery approach for Guard-Stage-Milestone models is proposed in \cite{DBLP:conf/bpm/PopovaD12,DBLP:journals/ijcis/PopovaFD15}, with an initial discovery of Petri nets from the event data and its conversion
to GSM models.

The discovery of composite state machines is introduced in \cite{DBLP:conf/bpm/EckSA18}, along with an implementation as a plug-in of the ProM framework and an assessment on the popular BPI Challenge 2017 dataset.
A refinement is proposed in \cite{DBLP:journals/bise/EckSA19}, allowing the enhancement of the model with performance metrics.

A user-guided approach for artifact-centric process modeling is proposed in \cite{DBLP:conf/enase/NguyenTO22} (BAPE, bottom-up artifact-centric process environments) with a process discovery integration
that allows to construct the overall process model from the event data of the fragments.

The discovery of object-aware processes in the PHILharmonicFlows platform is described in \cite{DBLP:conf/zeus/BreitmayerR20}.

{\bf ET2ET+ET2OT}: the basic idea of mainstream object-centric process discovery algorithms is to flatten the object-centric event log into the single object types,
apply a traditional process discovery algorithm for such logs and then collate the results together. This has been done in \cite{DBLP:conf/simpda/BertiA19,DBLP:conf/coopis/EldinAKBG22} (collation of directly-follows graphs)
and \cite{DBLP:journals/fuin/AalstB20} (collation of Petri nets). In \cite{DBLP:journals/sttt/BertiA23}, several new metrics (number of events, unique objects, total objects)
are proposed to decorate the diagrams. Also, MPM process mining proposes a model as a collation of models for the single object types \cite{Meyer2021AssociativeIF}.
An alternative approach includes the event-to-object and object-to-object relationships \cite{DBLP:conf/bpm/Berti19}.
The tool \cite{DBLP:conf/apn/AdamsA22} offers object-centric process discovery capabilities (object-centric Petri nets; visualization of variants).

Case studies on the discovery of object-centric Petri nets from manufacturing systems are proposed in \cite{DBLP:conf/case/LugaresiM21,schuh2022methodology}.

A tool for the discovery of process cubes on top of object-centric event data is proposed in \cite{DBLP:journals/corr/abs-2202-05709}, which uses object-centric directly-follows graphs and
Petri nets.

{\bf ET2ET+OT2OT+ET2OT}: several papers have been published on the discovery of object-centric behavioral constraints
models \cite{DBLP:conf/bis/LiCA17} \cite{DBLP:journals/emisa/LiC18} \cite{DBLP:journals/access/XiuLL22}.
Based on the constraints of the object model, an event correlation approach is also defined \cite{DBLP:conf/icws/LiCA18}.
Also, some example applications of the discovery on top of event logs extracted from the Dollibar ERP system and from popular social media platforms \cite{DBLP:journals/access/LiC19c,DBLP:conf/icccnt/DasK20} are proposed.
Object-centric behavioral constraint models are very sensitive to noise. Therefore, \cite{DBLP:journals/access/XiuLL22}
introduces a technique to manage the noise during the discovery of such process models.
Tool support is available \cite{DBLP:journals/emisa/LiC18}, but it is not working with the current version of the ProM framework.

In \cite{DBLP:conf/icpm/AaliMT21}, the care pathways for multi-morbid patients have been visualized using event graphs.
These graphs include activity-to-activity, activity-to-entity, and entity-to-entity relationships.

In \cite{DBLP:books/sp/22/Fahland22}, a process discovery approach for \emph{event knowledge graphs} is proposed, in which the relationships between the single events and objects are represented. Interestingly, through querying and the construction of views, different process representations can be obtained, from object-centric directly-follows graphs to Proclets (without advanced synchronization mechanisms). 

Finally, \cite{} is, to the best of our knowledge, the first proposal to approach the discovery problem for Petri nets with identifiers. Specifically, \cite{DBLP:conf/apn/BarenholzMPRRW23} formally investigates the flattening approach used to deal with {\bf ET2ET+ET2OT}, showing that it works for a class of Petri nets with identifiers, in the sense that it preserves a key re-discoverability property.

\subsubsection{Conformance Checking Techniques}

Here, we divide between techniques performing conformance checking on models satisfying {\bf ET2ET+OT2OT} (but not {\bf ET2OT}) or {\bf ET2ET+ET2OT} (but not {\bf OT2OT}).

{\bf ET2ET+OT2OT}: the main idea to perform conformance checking, proposed in \cite{DBLP:conf/bpm/FahlandLDA11,DBLP:conf/bis/FahlandLDA11},
is to split the problem into \emph{behavioral conformance} of the single entities (this can be done using traditional conformance checking techniques such as token-based replay and alignments)
and \emph{interaction conformance} between different entities.

The approach proposed in \cite{DBLP:journals/sosym/EstanolMCT19} is based on checking different conditions (constraints over the maximum number of related entities;
conditions over the transitions in the lifecycles) starting from a BAUML model.

In \cite{DBLP:conf/rcis/Breitmayer0R22}, conformance checking over
object-aware processes is done, with a formalism that allows for flexible lifecycle executions (different workflow nets are created for different conformance categories).

{\bf ET2ET+ET2OT}: Object-centric Petri nets have been proposed in \cite{DBLP:journals/fuin/AalstB20}, which allow for conformance checking on object-centric event logs.
On top of object-centric Petri nets, the concepts of fitness and precision have been defined \cite{DBLP:conf/icpm/AdamsA21}.

In \cite{DBLP:conf/icpm/ParkA22}, \emph{object-centric constraint graphs} are introduced with the purpose of representing constraints considering the interaction of objects.
These constraints are checked on a live stream of events.

A tool support offering support for object-centric conformance checking is described in \cite{DBLP:journals/sttt/BertiA23}.
This allows for the application of standard conformance checking techniques on the flattened logs (e.g., the log skeleton and the temporal profile), but also conformance rules based on the interaction
between different objects in the object-centric event log.
The proprietary tool of \cite{Meyer2021AssociativeIF} also implements an object-centric conformance checking approach, but the exact specification is unavailable.

\subsubsection{Performance Analysis Techniques}

Here, we collect some performance analysis techniques on top of object-centric process models.

Among models satisfying {\bf ET2ET+OT2OT}, but not {\bf ET2OT}, \cite{DBLP:journals/bise/EckSA19} allows for the decoration of composite state machine models with performance metrics (are delays in the processing of an artifact
correlated with delays in another artifact?).
Among models satisfying {\bf ET2ET+ET2OT}, but not {\bf OT2OT}, we cite the tool OC-PM \cite{DBLP:journals/sttt/BertiA23}.
Moreover, \cite{DBLP:conf/er/ParkAA22} introduces the enhancement of object-centric Petri nets with performance metrics.

Predictive analytics techniques are described in \cite{DBLP:journals/eswa/GalantiLNM23,DBLP:conf/icsoc/AdamsPLSA22,DBLP:conf/icsoc/GherissiHG22}.
The prediction of variables such as the total throughput time is possible thanks to learning a model out of the features of the objects of the object-centric event log and their interactions.

\begin{table*}
\caption{Summary of the included results.}
\centering
\resizebox{0.9\textwidth}{!}{
\begin{tabular}{|p{8cm}|p{6cm}|c|ccc|ccc|cccccc|cccc|}
\hline
\multicolumn{3}{|c|}{~} & \multicolumn{3}{|c|}{Log Type} & \multicolumn{3}{|c|}{Model Type} & \multicolumn{6}{|c|}{Technique} & \multicolumn{4}{|c|}{Quality} \\
\rotatebox[origin=c]{90}{{\bf Title}} & \rotatebox[origin=c]{90}{{\bf Authors}} & \rotatebox[origin=c]{90}{{\bf Year-Month}} & \rotatebox[origin=c]{90}{{\bf E2O}} & \rotatebox[origin=c]{90}{{\bf O2O}} & \rotatebox[origin=c]{90}{{\bf E2E}} & \rotatebox[origin=c]{90}{{\bf ET2ET}} & \rotatebox[origin=c]{90}{{\bf OT2OT}} & \rotatebox[origin=c]{90}{{\bf ET2OT}} & \rotatebox[origin=c]{90}{{\bf Extraction}} & \rotatebox[origin=c]{90}{{\bf Storage}} & \rotatebox[origin=c]{90}{{\bf Preprocessing}} & \rotatebox[origin=c]{90}{{\bf Discovery}} & \rotatebox[origin=c]{90}{{\bf Conformance}} & \rotatebox[origin=c]{90}{{\bf Perf.Analysis}} & \rotatebox[origin=c]{90}{{\bf Scalability}} & \rotatebox[origin=c]{90}{{\bf Generalization}} & \rotatebox[origin=c]{90}{{\bf Tool Support}} & \rotatebox[origin=c]{90}{{\bf Case Study}} \\
\hline
A Virtual Knowledge Graph Based Approach for Object-Centric Event Logs Extraction \cite{DBLP:conf/icpm/Xiong0KMGC22} & {\tiny Jing Xiong and Guohui Xiao and Tahir Emre Kalayci and Marco Montali and Zhenzhen Gu and Diego Calvanese} & 202310 & X & ~ & ~ & ~ & ~ & ~ & X & ~ & ~ & ~ & ~ & ~ & ~ & X & X & ~ \\
\hline
Process mining for artifact-centric blockchain applications \cite{m2023process} & {\tiny Leyla Moctar Mbaba and Nour Assy and Mohamed Sellami and Walid Gaaloul and Mohamedade Farouk Nanne} & 202309 & ~ & ~ & ~ & ~ & ~ & ~ & X & X & ~ & ~ & ~ & ~ & ~ & ~ & ~ & X \\
\hline
Normalizing object-centric process logs by applying database principles \cite{DBLP:journals/is/KumarST23} & {\tiny Akhil Kumar and Pnina Soffer and Arava Tsoury} & 202305 & X & ~ & ~ & ~ & ~ & ~ & ~ & X & X & ~ & ~ & ~ & ~ & X & ~ & ~ \\
\hline
Object-centric Process Predictive Analytics \cite{DBLP:journals/eswa/GalantiLNM23} & {\tiny Riccardo Galanti and Massimiliano de Leoni and Nicol{\`{o}} Navarin and Alan Marazzi} & 202303 & X & ~ & ~ & ~ & ~ & ~ & ~ & ~ & ~ & ~ & ~ & X & ~ & X & ~ & ~ \\
\hline
Object Type Clustering Using Markov Directly-Follow Multigraph in Object-Centric Process Mining \cite{DBLP:journals/access/Jalali22} & {\tiny Amin Jalali} & 202212 & X & ~ & ~ & ~ & ~ & ~ & ~ & ~ & X & ~ & ~ & ~ & ~ & X & ~ & ~ \\
\hline
A Framework for Extracting and Encoding Features from Object-Centric Event Data \cite{DBLP:conf/icsoc/AdamsPLSA22} & {\tiny Jan Niklas Adams and Gyunam Park and Sergej Levich and Daniel Schuster and Wil M. P. van der Aalst} & 202211 & X & ~ & ~ & ~ & ~ & ~ & ~ & ~ & X & ~ & ~ & X & ~ & X & X & ~ \\
\hline
Object-Centric Predictive Process Monitoring \cite{DBLP:conf/icsoc/GherissiHG22} & {\tiny Wissam Gherissi and Joyce El Haddad and Daniela Grigori} & 202211 & X & ~ & ~ & ~ & ~ & ~ & ~ & ~ & X & ~ & ~ & X & ~ & X & ~ & ~ \\
\hline
Enabling Multi-process Discovery on Graph Databases \cite{DBLP:conf/coopis/EldinAKBG22} & {\tiny Ali Nour Eldin and Nour Assy and Meriana Kobeissi and Jonathan Baudot and Walid Gaaloul} & 202210 & X & ~ & ~ & X & ~ & X & ~ & ~ & ~ & X & ~ & ~ & X & X & ~ & ~ \\
\hline
OPerA: Object-Centric Performance Analysis \cite{DBLP:conf/er/ParkAA22} & {\tiny Gyunam Park and Jan Niklas Adams and Wil M. P. van der Aalst} & 202210 & X & ~ & ~ & X & ~ & X & ~ & ~ & ~ & X & ~ & X & ~ & X & X & ~ \\
\hline
ocpa: A Python library for object-centric process analysis \cite{adams2022ocpa} & {\tiny Jan Niklas Adams and Gyunam Park and Wil M. P. van der Aalst} & 202210 & X & ~ & ~ & X & ~ & X & ~ & X & X & X & X & X & ~ & X & X & ~ \\
\hline
Enhancing Data-Awareness of Object-Centric Event Logs \cite{DBLP:conf/icpm/GoossensSVA22} & {\tiny Alexandre Goossens and Johannes De Smedt and Jan Vanthienen and Wil M. P. van der Aalst} & 202210 & X & ~ & ~ & ~ & ~ & ~ & ~ & X & X & ~ & ~ & ~ & ~ & X & ~ & ~ \\
\hline
Interactive Process Identification and Selection from SAP ERP \cite{DBLP:conf/icpm/WeberPRA22} & {\tiny Julian Weber and Gyunam Park and Majid Rafiei and Wil M.P. van der Aalst} & 202210 & X & ~ & ~ & ~ & ~ & ~ & X & ~ & ~ & ~ & ~ & ~ & ~ & ~ & X & ~ \\
\hline
Konekti: A Data Preparation Platform for Process Mining \cite{DBLP:conf/icpm/VugsAS22} & {\tiny Lotte Vugs and Maarten van Asseldonk and Niek van Son} & 202210 & X & ~ & ~ & ~ & ~ & ~ & X & X & ~ & ~ & ~ & ~ & ~ & X & X & ~ \\
\hline
Clustering Analysis and Frequent Pattern Mining for Process Profile Analysis: An Exploratory Study for Object-Centric Event Logs \cite{faria2022clustering} & {\tiny Faria Junior and Elio Ribeiro and Thais Rodrigues Neubauer and Marcelo Fantinato and Sarajane Marques Peres} & 202210 & X & ~ & ~ & ~ & ~ & ~ & ~ & ~ & X & ~ & ~ & ~ & ~ & X & ~ & ~ \\
\hline
Defining Cases and Variants for Object-Centric Event Data \cite{DBLP:conf/icpm/AdamsSSSA22} & {\tiny Jan Niklas Adams and Daniel Schuster and Seth Schmitz and Gunther Schuh and Wil M. P. van der Aalst} & 202210 & X & ~ & ~ & ~ & ~ & ~ & ~ & ~ & X & ~ & ~ & ~ & ~ & X & X & ~ \\
\hline
Monitoring Constraints in Business Processes Using Object-Centric Constraint Graphs \cite{DBLP:conf/icpm/ParkA22} & {\tiny Gyunam Park and Wil M. P. van der Aalst} & 202210 & X & ~ & ~ & ~ & ~ & ~ & ~ & ~ & ~ & ~ & X & ~ & ~ & X & X & ~ \\
\hline
Uncovering Object-Centric Data in Classical Event Logs for the Automated Transformation from XES to OCEL \cite{DBLP:conf/bpm/RebmannRA22} & {\tiny Adrian Rebmann and Jana-Rebecca Rehse and Han van der Aa} & 202209 & X & ~ & ~ & ~ & ~ & ~ & ~ & X & X & ~ & ~ & ~ & ~ & X & ~ & ~ \\
\hline
Assessing the Suitability of Traditional Event Log Standards for IoT-Enhanced Event Logs \cite{DBLP:conf/bpm/BertrandWS22} & {\tiny Yannis Bertrand and Jochen De Weerdt and Estefania Serral} & 202209 & X & ~ & ~ & ~ & ~ & ~ & ~ & X & ~ & ~ & ~ & ~ & ~ & ~ & ~ & ~ \\
\hline
OC-PM: Analyzing Object-Centric Event Logs and Process Models \cite{DBLP:journals/sttt/BertiA23} & {\tiny Alessandro Berti and Wil M. P. van der Aalst} & 202208 & X & ~ & ~ & X & ~ & X & ~ & ~ & ~ & X & X & X & X & X & X & ~ \\
\hline
Discovery of Object-Centric Behavioral Constraint Models With Noise \cite{DBLP:journals/access/XiuLL22} & {\tiny Baoxin Xiu and Guangming Li and Yidan Li} & 202208 & X & X & ~ & X & X & X & ~ & ~ & ~ & X & ~ & ~ & ~ & X & X & ~ \\
\hline
Extraction of Object-Centric Event Logs through Virtual Knowledge Graphs \cite{DBLP:conf/dlog/Xiong0KMGC22} & {\tiny Jing Xiong and Guohui Xiao and Tahir Emre Kalayci and Marco Montali and Zhenzhen Gu and Diego Calvanese} & 202208 & X & ~ & ~ & X & ~ & X & X & ~ & ~ & ~ & ~ & ~ & ~ & X & X & ~ \\
\hline
Extracting Artifact-Centric Event Logs From Blockchain Applications \cite{DBLP:conf/IEEEscc/Moctar-MBabaASG22} & {\tiny Leyla Moctar-M'Baba and Nour Assy and Mohamed Sellami and Walid Gaaloul and Mohamedade Farouk Nanne} & 202207 & ~ & X & ~ & ~ & ~ & ~ & X & X & ~ & ~ & ~ & ~ & ~ & ~ & X & ~ \\
\hline
Petri net-based object-centric processes with read-only data \cite{DBLP:journals/is/GhilardiGMR22} & {\tiny Silvio Ghilardi and Alessandro Gianola and Marco Montali and Andrey Rivkin} & 202207 & ~ & ~ & ~ & ~ & ~ & ~ & ~ & ~ & ~ & ~ & ~ & ~ & ~ & X & ~ & ~ \\
\hline
OC-$\pi$: Object-Centric Process Insights \cite{DBLP:conf/apn/AdamsA22} & {\tiny Jan Niklas Adams and Wil M. P. van der Aalst} & 202206 & X & ~ & ~ & X & ~ & X & ~ & ~ & X & X & ~ & ~ & ~ & X & X & ~ \\
\hline
Process Mining over Multiple Behavioral Dimensions with Event Knowledge Graphs \cite{DBLP:books/sp/22/Fahland22} & {\tiny Dirk Fahland} & 202206 & X & X & X & ~ & ~ & ~ & ~ & ~ & ~ & X & ~ & ~ & ~ & X & ~ & ~ \\
\hline
Data and Process Resonance - Identifier Soundness for Models of Information Systems \cite{DBLP:conf/apn/WerfRPM22} & {\tiny Jan Martijn E. M. van der Werf and Andrey Rivkin and Artem Polyvyanyy and Marco Montali} & 202206 & ~ & ~ & ~ & ~ & ~ & ~ & ~ & ~ & ~ & ~ & ~ & ~ & ~ & X & ~ & ~ \\
\hline
Enabling Conformance Checking for Object Lifecycle Processes \cite{DBLP:conf/rcis/Breitmayer0R22} & {\tiny Marius Breitmayer and Lisa Arnold and Manfred Reichert} & 202205 & ~ & X & ~ & X & X & ~ & ~ & ~ & ~ & ~ & X & ~ & ~ & ~ & X & ~ \\
\hline
Process Mining to Discover the Global Process from its Fragments' Executions \cite{DBLP:conf/enase/NguyenTO22} & {\tiny Minh Khoi Nguyen and Hanh Nhi Tran and Ileana Ober} & 202204 & ~ & X & ~ & X & X & ~ & ~ & ~ & X & X & ~ & ~ & ~ & X & ~ & ~ \\
\hline
A Methodology to Apply Process Mining in End-To-End Order Processing of Manufacturing Companies \cite{schuh2022methodology} & {\tiny Schuh G. and Gutzlaff A. and Schmitz S. and Kuhn C. and Klapper N.} & 202201 & X & ~ & ~ & X & ~ & X & ~ & ~ & X & X & ~ & ~ & ~ & ~ & ~ & X \\
\hline
Multi-Dimensional Event Data in Graph Databases \cite{DBLP:journals/jodsn/EsserF21} & {\tiny Stefan Esser and Dirk Fahland} & 202112 & X & X & X & X & X & X & ~ & X & X & X & ~ & ~ & X & X & X & ~ \\
\hline
An Event Data Extraction Approach from SAP ERP for Process Mining \cite{DBLP:conf/icpm/BertiPRA21} & {\tiny Alessandro Berti and Gyunam Park and Majid Rafiei and Wil M. P. van der Aalst} & 202111 & X & ~ & ~ & ~ & ~ & ~ & X & ~ & X & ~ & ~ & ~ & ~ & X & X & ~ \\
\hline
Associative Intelligence for Object-Centric Process Mining with MPM \cite{Meyer2021AssociativeIF} & {\tiny Janna Meyer and Josua Reimold} & 202111 & X & ~ & ~ & X & ~ & X & ~ & ~ & ~ & X & X & ~ & X & ~ & X & ~ \\
\hline
A Scalable Database for the Storage of Object-Centric Event Logs \cite{DBLP:journals/corr/abs-2202-05639} & {\tiny Alessandro Berti and Anahita Farhang Ghahfarokhi and Gyunam Park and Wil M. P. van der Aalst} & 202111 & X & ~ & ~ & ~ & ~ & ~ & ~ & X & X & X & ~ & ~ & X & X & X & ~ \\
\hline
Precision and Fitness in Object-Centric Process Mining \cite{DBLP:conf/icpm/AdamsA21} & {\tiny Jan Niklas Adams and Wil M. P. van der Aalst} & 202111 & X & ~ & ~ & X & ~ & X & ~ & ~ & ~ & ~ & X & ~ & ~ & X & ~ & ~ \\
\hline
A Python Tool for Object-Centric Process Mining Comparison \cite{DBLP:journals/corr/abs-2202-05709} & {\tiny Anahita Farhang Ghahfarokhi and Wil M. P. van der Aalst} & 202111 & X & ~ & ~ & X & ~ & X & ~ & ~ & X & X & ~ & X & ~ & X & X & ~ \\
\hline
Discovering Care Pathways for Multi-morbid Patients Using Event Graphs \cite{DBLP:conf/icpm/AaliMT21} & {\tiny Milad Naeimaei Aali and Felix Mannhardt and Pieter Jelle Toussaint} & 202110 & X & X & ~ & X & X & X & ~ & ~ & ~ & X & ~ & ~ & ~ & ~ & ~ & ~ \\
\hline
Federated Process Mining: Exploiting Event Data Across Organizational Boundaries \cite{DBLP:conf/smds/Aalst21} & {\tiny Wil M. P. van der Aalst} & 202109 & ~ & X & X & ~ & ~ & ~ & ~ & ~ & X & ~ & ~ & ~ & ~ & X & ~ & ~ \\
\hline
Storing and Querying Multi-dimensional Process Event Logs Using Graph Databases \cite{DBLP:conf/bpm/EsserF19} & {\tiny Stefan Esser and Dirk Fahland} & 202109 & X & X & X & X & X & X & ~ & X & X & X & ~ & ~ & X & X & X & ~ \\
\hline
OCEL: A Standard for Object-Centric Event Logs \cite{DBLP:conf/adbis/GhahfarokhiPBA21} & {\tiny Anahita Farhang Ghahfarokhi and Gyunam Park and Alessandro Berti and Wil M. P. van der Aalst} & 202108 & X & ~ & ~ & ~ & ~ & ~ & ~ & X & ~ & ~ & ~ & ~ & ~ & X & X & ~ \\
\hline
Discovery and digital model generation for manufacturing systems with assembly operations \cite{DBLP:conf/case/LugaresiM21} & {\tiny Giovanni Lugaresi and Andrea Matta} & 202108 & X & ~ & ~ & X & ~ & X & ~ & ~ & X & X & ~ & ~ & ~ & ~ & ~ & X \\
\hline
Towards a Domain-Specific Modeling Language for Extracting Event Logs from ERP Systems \cite{pajic2021towards} & {\tiny Pajic Simovic Ana and Babarogic Sladjan and Pantelic Ognjen and Krstovic Stefan} & 202106 & ~ & X & ~ & ~ & ~ & ~ & X & ~ & ~ & ~ & ~ & ~ & ~ & ~ & ~ & ~ \\
\hline
Merging event logs for inter-organizational process mining \cite{hernandez2021merging} & {\tiny Hernandez-Resendiz, Jaciel David and Tello-Leal, Edgar and Marin-Castro, Heidy Marisol and Ramirez-Alcocer, Ulises Manuel and Mata-Torres, Jonathan Alfonso} & 202106 & ~ & X & X & ~ & ~ & ~ & ~ & ~ & X & ~ & ~ & ~ & ~ & X & X & ~ \\
\hline
Discovering Object-centric Petri Nets \cite{DBLP:journals/fuin/AalstB20} & {\tiny Wil M. P. van der Aalst and Alessandro Berti} & 202012 & X & ~ & ~ & X & ~ & X & ~ & ~ & ~ & X & X & X & ~ & X & X & ~ \\
\hline
Case notion discovery and recommendation: automated event log building on databases \cite{DBLP:journals/kais/MurillasRA20} & {\tiny Eduardo Gonzalez Lopez de Murillas and Hajo A. Reijers and Wil M. P. van der Aalst} & 202007 & ~ & ~ & ~ & ~ & ~ & ~ & X & X & X & ~ & ~ & ~ & ~ & X & X & ~ \\
\hline
Behavioural Analysis of Multi-Source Social Network Data Using Object-Centric Behavioural Constraints and Data Mining Technique \cite{DBLP:conf/icccnt/DasK20} & {\tiny Priyam Das and Dhananjay R. Kalbande} & 202007 & X & X & ~ & X & X & X & X & ~ & ~ & X & ~ & ~ & ~ & ~ & ~ & ~ \\
\hline
\end{tabular}
}
\label{tab:includedResults}
\end{table*}
\begin{table*}
\centering
\resizebox{0.9\textwidth}{!}{
\begin{tabular}{|p{8cm}|p{6cm}|c|ccc|ccc|cccccc|cccc|}
\hline
\multicolumn{3}{|c|}{~} & \multicolumn{3}{|c|}{Log Type} & \multicolumn{3}{|c|}{Model Type} & \multicolumn{6}{|c|}{Technique} & \multicolumn{4}{|c|}{Quality} \\
\rotatebox[origin=c]{90}{{\bf Title}} & \rotatebox[origin=c]{90}{{\bf Authors}} & \rotatebox[origin=c]{90}{{\bf Year-Month}} & \rotatebox[origin=c]{90}{{\bf E2O}} & \rotatebox[origin=c]{90}{{\bf O2O}} & \rotatebox[origin=c]{90}{{\bf E2E}} & \rotatebox[origin=c]{90}{{\bf ET2ET}} & \rotatebox[origin=c]{90}{{\bf OT2OT}} & \rotatebox[origin=c]{90}{{\bf ET2OT}} & \rotatebox[origin=c]{90}{{\bf Extraction}} & \rotatebox[origin=c]{90}{{\bf Storage}} & \rotatebox[origin=c]{90}{{\bf Preprocessing}} & \rotatebox[origin=c]{90}{{\bf Discovery}} & \rotatebox[origin=c]{90}{{\bf Conformance}} & \rotatebox[origin=c]{90}{{\bf Perf.Analysis}} & \rotatebox[origin=c]{90}{{\bf Scalability}} & \rotatebox[origin=c]{90}{{\bf Generalization}} & \rotatebox[origin=c]{90}{{\bf Tool Support}} & \rotatebox[origin=c]{90}{{\bf Case Study}} \\
\hline
Towards the discovery of object-aware processes \cite{DBLP:conf/zeus/BreitmayerR20} & {\tiny Marius Breitmayer and Manfred Reichert} & 202002 & ~ & X & ~ & X & X & ~ & ~ & ~ & ~ & X & ~ & ~ & ~ & ~ & X & ~ \\
\hline
Guided Interaction Exploration and Performance Analysis in Artifact-Centric Process Models \cite{DBLP:journals/bise/EckSA19} & {\tiny Maikel L. van Eck and Natalia Sidorova and Wil M. P. van der Aalst} & 201912 & ~ & X & ~ & X & X & ~ & ~ & ~ & ~ & X & ~ & X & ~ & X & X & ~ \\
\hline
Object-Centric Process Mining: Dealing with Divergence and Convergence in Event Data \cite{DBLP:conf/sefm/Aalst19} & {\tiny Wil M. P. van der Aalst} & 201909 & X & ~ & ~ & ~ & ~ & ~ & ~ & ~ & ~ & X & ~ & ~ & ~ & X & ~ & ~ \\
\hline
Process Mining on Event Graphs: a Framework to Extensively Support Projects \cite{DBLP:conf/bpm/Berti19} & {\tiny Alessandro Berti} & 201909 & X & X & X & ~ & ~ & ~ & ~ & ~ & X & ~ & ~ & ~ & ~ & X & ~ & ~ \\
\hline
Extracting Multiple Viewpoint Models from Relational Databases \cite{DBLP:conf/simpda/BertiA19} & {\tiny Alessandro Berti and Wil M. P. van der Aalst} & 201908 & X & ~ & ~ & X & ~ & X & X & X & X & X & ~ & ~ & X & X & X & ~ \\
\hline
Process Mining in Social Media: Applying Object-Centric Behavioral Constraint Models \cite{DBLP:journals/access/LiC19c} & {\tiny Guangming Li and Renata Medeiros de Carvalho} & 201906 & X & X & ~ & X & X & X & X & ~ & ~ & X & ~ & ~ & ~ & ~ & ~ & ~ \\
\hline
Conformance checking in UML artifact-centric business process models \cite{DBLP:journals/sosym/EstanolMCT19} & {\tiny Montserrat Estanol and Jorge Munoz-Gama and Josep Carmona and Ernest Teniente} & 201902 & ~ & X & ~ & X & X & ~ & ~ & X & ~ & ~ & X & ~ & ~ & ~ & ~ & ~ \\
\hline
Dealing with Artifact-Centric Systems: a Process Mining Approach \cite{DBLP:journals/emisa/LiC18} & {\tiny Guangming Li and Renata Medeiros de Carvalho} & 201812 & X & X & ~ & X & X & X & X & X & ~ & X & ~ & ~ & ~ & ~ & X & ~ \\
\hline
Multi-instance Mining: Discovering Synchronisation in Artifact-Centric Processes \cite{DBLP:conf/bpm/EckSA18} & {\tiny Maikel L. van Eck and Natalia Sidorova and Wil M. P. van der Aalst} & 201809 & ~ & X & ~ & X & X & ~ & ~ & ~ & ~ & X & ~ & ~ & ~ & X & X & ~ \\
\hline
A domain-specific language for supporting event log extraction from ERP systems \cite{simovic2018domain} & {\tiny Simovic Ana Pajic and Babarogic Sladan and Pantelic Ognjen} & 201807 & ~ & X & ~ & ~ & ~ & ~ & X & X & ~ & ~ & ~ & ~ & ~ & ~ & ~ & ~ \\
\hline
Configurable Event Correlation for Process Discovery from Object-Centric Event Data \cite{DBLP:conf/icws/LiCA18} & {\tiny Guangming Li and Renata Medeiros de Carvalho and Wil M. P. van der Aalst} & 201807 & X & X & ~ & X & X & X & ~ & ~ & X & ~ & ~ & ~ & ~ & X & X & ~ \\
\hline
Extracting Object-Centric Event Logs to Support Process Mining on Databases \cite{DBLP:conf/caise/LiMCA18} & {\tiny Guangming Li and Eduardo Gonzalez Lopez de Murillas and Renata Medeiros de Carvalho and Wil M. P. van der Aalst} & 201806 & X & X & ~ & X & X & X & X & X & ~ & ~ & ~ & ~ & ~ & X & ~ & ~ \\
\hline
Merging event logs: Combining granularity levels for process flow analysis \cite{DBLP:journals/is/RaichelsonSV17} & {\tiny Lihi Raichelson and Pnina Soffer and Eric Verbeek} & 201711 & ~ & X & X & ~ & ~ & ~ & ~ & ~ & X & ~ & ~ & ~ & ~ & X & X & ~ \\
\hline
Automatic Discovery of Object-Centric Behavioral Constraint Models \cite{DBLP:conf/bis/LiCA17} & {\tiny Guangming Li and Renata Medeiros de Carvalho and Wil M. P. van der Aalst} & 201706 & X & X & ~ & X & X & X & X & ~ & ~ & X & ~ & ~ & ~ & X & X & ~ \\
\hline
Metamodel of the Artifact-Centric Approach to Event Log Extraction from ERP Systems \cite{DBLP:journals/ijdsst/PajicB16} & {\tiny Ana Pajic and Dragana Becejski-Vujaklija} & 201610 & ~ & X & ~ & ~ & ~ & ~ & X & ~ & ~ & ~ & ~ & ~ & ~ & ~ & ~ & ~ \\
\hline
Merging event logs for process mining with hybrid artificial immune algorithm \cite{xu2016merging} & {\tiny Xu Yang and Lin Qi and Zhao Martin Q.} & 201601 & ~ & X & X & ~ & ~ & ~ & ~ & ~ & X & ~ & ~ & ~ & ~ & X & ~ & ~ \\
\hline
Discovering Interacting Artifacts from ERP Systems \cite{DBLP:journals/tsc/LuNWF15} & {\tiny Xixi Lu and Marijn Nagelkerke and Dennis van de Wiel and Dirk Fahland} & 201508 & ~ & X & ~ & X & X & ~ & X & ~ & X & X & ~ & ~ & ~ & ~ & X & ~ \\
\hline
Process Mining on Databases: Unearthing Historical Data from Redo Logs \cite{DBLP:conf/bpm/MurillasAR15} & {\tiny Eduardo Gonzalez Lopez de Murillas and Wil M. P. van der Aalst and Hajo A. Reijers} & 201508 & ~ & ~ & ~ & ~ & ~ & ~ & X & X & X & ~ & ~ & ~ & ~ & X & X & ~ \\
\hline
Merging Event Logs with Many to Many Relationships \cite{DBLP:conf/bpm/RaichelsonS14} & {\tiny Lihi Raichelson and Pnina Soffer} & 201409 & ~ & X & X & ~ & ~ & ~ & ~ & ~ & X & ~ & ~ & ~ & ~ & X & ~ & ~ \\
\hline
Merging event logs for process mining: A rule based merging method and rule suggestion algorithm \cite{DBLP:journals/eswa/ClaesP14} & {\tiny Jan Claes and Geert Poels} & 201407 & ~ & X & X & ~ & ~ & ~ & ~ & ~ & X & ~ & ~ & ~ & ~ & X & X & ~ \\
\hline
Discovering Unbounded Synchronization Conditions in Artifact-Centric Process Models \cite{DBLP:conf/bpm/PopovaD13} & {\tiny Niels Lohmann and Minseok Song and Petia Wohed} & 201308 & ~ & X & ~ & X & X & ~ & ~ & ~ & ~ & X & ~ & ~ & ~ & ~ & ~ & ~ \\
\hline
Artifact Lifecycle Discovery \cite{DBLP:journals/ijcis/PopovaFD15} & {\tiny Viara Popova and Dirk Fahland and Marlon Dumas} & 201303 & ~ & X & ~ & X & X & ~ & ~ & ~ & ~ & X & ~ & ~ & ~ & ~ & ~ & ~ \\
\hline
Automatic Discovery of Data-Centric and Artifact-Centric Processes \cite{DBLP:conf/bpm/NooijenDF12} & {\tiny Erik H. J. Nooijen and Boudewijn F. van Dongen and Dirk Fahland} & 201209 & ~ & X & ~ & X & X & ~ & X & ~ & ~ & X & ~ & ~ & ~ & ~ & ~ & ~ \\
\hline
From Petri Nets to Guard-Stage-Milestone Models \cite{DBLP:conf/bpm/PopovaD12} & {\tiny Viara Popova and Marlon Dumas} & 201209 & ~ & X & ~ & X & X & ~ & ~ & ~ & ~ & X & ~ & ~ & ~ & ~ & ~ & ~ \\
\hline
Conformance Checking of Interacting Processes with Overlapping Instances \cite{DBLP:conf/bpm/FahlandLDA11} & {\tiny Dirk Fahland and Massimiliano de Leoni and Boudewijn F. van Dongen and Wil M. P. van der Aalst} & 201109 & ~ & X & ~ & X & X & ~ & ~ & ~ & ~ & ~ & X & ~ & ~ & X & ~ & ~ \\
\hline
Behavioral Conformance of Artifact-Centric Process Models \cite{DBLP:conf/bis/FahlandLDA11} & {\tiny Dirk Fahland and Massimiliano de Leoni and Boudewijn F. van Dongen and Wil M. P. van der Aalst} & 201106 & ~ & X & ~ & X & X & ~ & ~ & ~ & ~ & ~ & X & ~ & ~ & X & ~ & ~ \\
\hline
\end{tabular}
}
\end{table*}

\begin{figure}[ht]
\centering
\includegraphics[width=\textwidth]{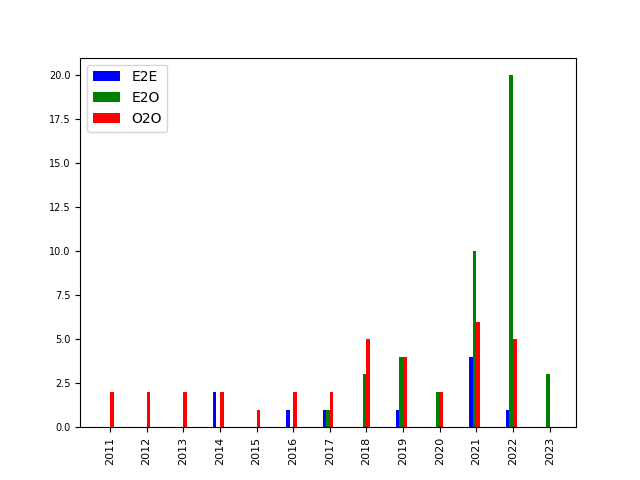}
\caption{Categories of event logs used in the considered papers.}
\label{fig:typesEventLogsMatplotlib}
\end{figure}

\begin{figure}[ht]
\centering
\includegraphics[width=\textwidth]{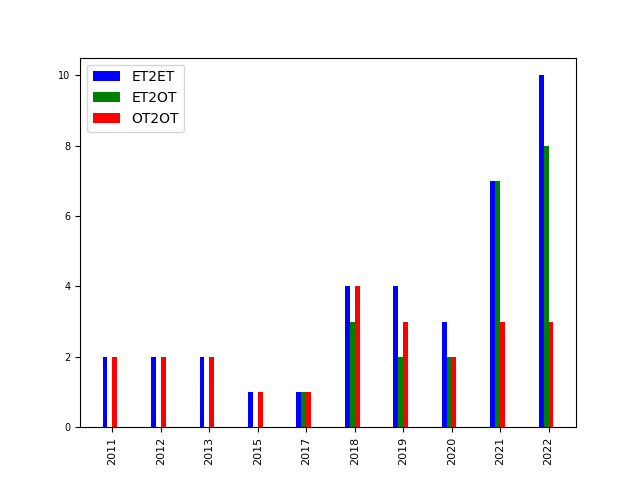}
\caption{Categories of process models used in the considered papers.}
\label{fig:typesProcessModelsMatplotlib}
\end{figure}

\begin{figure}[ht]
\centering
\includegraphics[width=\textwidth]{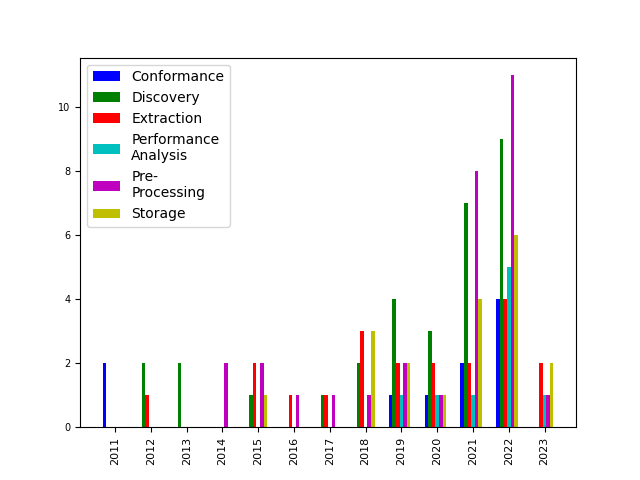}
\caption{Categories of techniques used in the considered papers.}
\label{fig:techniquesMatplotlib}
\end{figure}

\subsection{Dimensions}
\label{subsec:paperDimensions}

In this section, we verify the dimensions of the papers under different criteria (scalability, generalization, availability of tool support,
and case studies on real-life event data).
We divide the analysis between techniques working with process models belonging to the {\bf ET2ET} and the {\bf ET2OT} categories (but not {\bf OT2OT}),
techniques working with process models belonging to the {\bf ET2ET} and the {\bf OT2OT} categories (but not {\bf ET2OT}); techniques working with process models belonging to the three categories
{\bf ET2ET}, {\bf OT2OT} and {\bf ET2OT}.

\begin{figure}[ht]
\centering
\includegraphics[width=\textwidth]{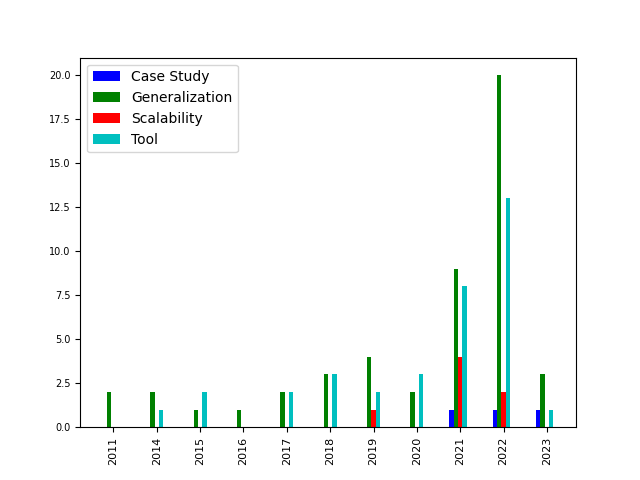}
\caption{Criteria used in the considered papers.}
\label{fig:qualityMatplotlib}
\end{figure}

\subsubsection{Scalability}

In this section, we discuss the scalability of the techniques as a dimension influencing the applicability of object-centric techniques in real-life contexts.

{\bf ET2ET+ET2OT}: since mainstream object-centric process discovery techniques are based on: i) flattening the object-centric event log on the different object types; ii) discovering a traditional process model on the flattened log;
iii) collating everything together, the discovery techniques (\cite{DBLP:conf/simpda/BertiA19,DBLP:journals/fuin/AalstB20,DBLP:journals/sttt/BertiA23})
have good scalability.
However, the process enhancement phase (decorating the model with frequency or performance metrics) requires a replay operation that has lower scalability
(considering the metrics described in \cite{DBLP:journals/fuin/AalstB20} and \cite{DBLP:conf/er/ParkAA22}).

{\bf ET2ET+OT2OT}: these models can be discovered from the event data in three different phases: i) discovery of the lifecycle of the entities;
ii) discovery of the interaction between different entities; iii) for some models, the discovery of the synchronization conditions between the entities.
While for the first phase process discovery algorithms on traditional event logs can be used, and good scalability could be achieved,
for the second and especially the third phase the scalability is generally problematic.

{\bf ET2ET+OT2OT+ET2OT}: the discovery of the declarative constraints of object-centric behavioral constraint models \cite{DBLP:conf/bis/LiCA17} scales poorly. Moreover, also the event log format (XOC) proposed in \cite{DBLP:conf/caise/LiMCA18}
is of difficult usage in real-life contexts due to the persistence of the status of the database for every event.

\subsubsection{Generalization}

In this section, we discuss the generalization of the techniques as a dimension influencing the applicability in different real-life scenarios.

{\bf ET2ET+ET2OT}: these techniques can be applied to object-centric event logs, and some publicly available object-centric event logs can be downloaded from \url{http://www.ocel-standard.org/}. Hence, most of these techniques are generic
and do not depend on the target information systems.

{\bf ET2ET+OT2OT}: composite state machines can be discovered starting from XES event logs using the \emph{CSM miner} plugin \cite{DBLP:conf/bpm/EckSA18}.
Therefore, the technique can be applied to any event log. Other techniques are highly dependent on the database (\cite{DBLP:conf/bpm/NooijenDF12})/information system (\cite{DBLP:journals/tsc/LuNWF15}),
hence they cannot be generalized.

{\bf ET2ET+OT2OT+ET2OT}: object-centric behavioral constraint techniques \cite{DBLP:conf/bis/LiCA17} start from XOC logs \cite{DBLP:conf/caise/LiMCA18}, hence they are generic. However, due to the problems in the scalability
and the lack of publicly available XOC logs, the application of such techniques to mainstream systems is difficult. The authors describe in a scientific paper the extraction of
event logs from the Dollibar ERP system \cite{DBLP:conf/caise/LiMCA18} and from popular social media platforms \cite{DBLP:journals/access/LiC19c}.

\subsubsection{Availability of Tool Support}

Two prominent tools in the object-centric field are OC-PM and OC-$\pi$.
OC-PM, as detailed in \cite{DBLP:journals/sttt/BertiA23}, provides a wide range of capabilities such as process discovery, conformance checking, machine learning, and filtering algorithms. On the other hand, OC-$\pi$, discussed in \cite{DBLP:conf/apn/AdamsA22}, specializes in process discovery and variants visualization.

Additional tools have also been developed to support object-centric process mining analyses, such as:

\begin{itemize}
\item The 'Process cubes' tool, according to \cite{DBLP:journals/corr/abs-2202-05709}, which is designed for process comparison purposes.
\item A proprietary tool outlined in \cite{Meyer2021AssociativeIF} that is specifically created for object-centric process mining.
\item A tool discussed in \cite{DBLP:conf/icpm/BertiPRA21}, which extracts OCELs from SAP ERP systems.
\item A Python library for object-centric process mining, as mentioned in \cite{adams2022ocpa}.
\end{itemize}

Moreover, there is tool support for the discovery of object-centric behavioral constraint models, as discussed in \cite{DBLP:conf/bis/LiCA17}. This tool exists as a plugin for the ProM framework, although its usage can be challenging and obtaining publicly available XOC logs can be difficult.

Lastly, while other models can be discovered, as demonstrated in \cite{DBLP:journals/tsc/LuNWF15}, the associated tool support is somewhat dated and can be challenging to initiate.

\subsubsection{Case Studies using Real-Life Event Data}

Here, we analyze which papers included in the review have applied their techniques to case studies on real-life information systems.
These would help to prove the readiness of the discipline. This criterion excludes papers with an assessment done on a publicly available real-life event log,
because the questions and the results have not been discussed with the business.

Unfortunately, only a few case studies have been proposed.
In \cite{DBLP:conf/case/LugaresiM21}, the problem of modeling a production system including different assembly stations (where every station records the event data singularly)
using object-centric process models is assessed, and a flow shop including the assembly stations is successfully discovered. This allows for building a simulation model with the desired behavior.
In \cite{schuh2022methodology}, an end-to-end object-centric process is extracted with different stages of the manufacturing process (sales, manufacturing, shipping).

In \cite{m2023process}, an application of artifact-centric process mining to public Ethereum applications is proposed.

\subsection{Temporal Evolution of the Discipline}
\label{subsec:temporalEvolution}

The discipline of object-centric process mining has experienced a fascinating trajectory over the past years, which can be charted in terms of various aspects. Four specific viewpoints have been instrumental in illustrating this temporal evolution.

Primarily, the usage of different categories of event logs has undergone a noteworthy change. As depicted in Fig. \ref{fig:typesEventLogsMatplotlib}, event logs containing object-to-object relationships were predominantly utilized between 2011 and 2016. However, more recently, there has been a discernible shift towards event logs where a single event can correspond to multiple objects of various object types.

Secondly, the choice of process models during the same timeframe presents a similar pattern. As shown in Fig. \ref{fig:typesProcessModelsMatplotlib}, between 2011 and 2016, models illustrating {\bf ET2ET+OT2OT} interactions were favored. In more recent years, however, the preference has steered towards process models that represent {\bf ET2ET+ET2OT} interactions.

The classification in techniques provides insights into the evolution of the discipline from a more operational perspective. From Fig. \ref{fig:techniquesMatplotlib}, it is observable that in recent times, greater emphasis has been laid on aspects of storage, preprocessing, and performance analysis. This underlines the increasing maturity of the field, moving from the creation and understanding of models to more practical aspects of process mining.

Finally, some criteria have been used to evaluate the progress in the discipline, as shown in Fig. \ref{fig:qualityMatplotlib}. This includes criteria like scalability, generalization, tool support, and case studies. Over the years, the availability of tool support has noticeably increased, reflecting the discipline's progression toward practical applicability. Additionally, there's a growing focus on scalability, signifying the rising interest in deploying these techniques in more extensive and complex environments.

In summary, the temporal evolution of object-centric process mining, as seen through these four viewpoints, paints a picture of a rapidly maturing discipline. It is transitioning from a primarily theoretical field with a focus on defining models and interactions to a more practical discipline where the emphasis is increasingly on applicability, scalability, and performance. Together with more scalable and practical techniques, we expect an increasing number of papers dealing with richer process models that support {\bf ET2ET+ET2OT+OT2OT} features and where discovery and conformance checking consider advanced forms of synchronization as a first-class citizen.

\section{Conclusion}
\label{sec:discussion}

As we conclude this comprehensive review, we briefly assess the current maturity of the object-centric process mining discipline and identify the challenges that remain to be addressed.

The discipline has been witnessing a steady surge in interest, as evinced by the increasing number of publications annually. This speaks volumes about the vitality and relevance of the field. Progress has been marked in terms of standard formats for storing object-centric event logs. The first milestone in this direction is constituted by the OCEL format, backed by numerous implementations on scalable databases. This advancement has helped facilitate the broader adoption and implementation of object-centric process mining techniques.

Recent years have also seen the advent of several preprocessing techniques. These are crucial for conducting real-life case studies of object-centric process mining. Similarly, the pool of process discovery techniques that start from object-centric event data has grown, with a few demonstrating a commendable degree of scalability. The interest in object-centric model enhancement, too, has been on the rise.

Another encouraging development is the emergence of open-source object-centric process mining tools that are publicly available. The OC-PM tool and OC-$\pi$ are notable examples. In tandem with this, some object-centric event logs are also accessible on the OCEL standard website. Further enriching the discipline are the real-life case studies of object-centric process mining that have been carried out over the last few years.

Despite these strides, some challenges persist. The limited scalability of many techniques, for instance, continues to be a roadblock. Other hurdles include the lack of support for extracting object-centric event data from mainstream systems like ERPs and CRMs, and the challenge of effectively laying out object-centric process models.

Finally, while many process models are essentially adaptations of traditional process models such as DFGs and Petri nets, a few innovative contributions are worth mentioning. These include the OCBC models, event knowledge graphs, Proclets, Petri nets with identifiers, and all the more advanced process models discussed in Section~\ref{subsec:models}. These models leverage more advanced constructs and interactions between different object types, opening up new avenues for exploration. However, enhancing the scalability of these innovative techniques remains a significant area of investigation.

In conclusion, while object-centric process mining has certainly come a long way, it is still a developing field with plenty of opportunities and challenges ahead.

\section*{Declarations}

{\bf Ethical Approval}: Not Applicable.

\noindent
{\bf Availability of supporting data}: Not Applicable.

\noindent
{\bf Competing Interests}: The authors declare that they have no known competing financial interests or personal relationships that could have appeared to influence the work reported in this paper.

\noindent
{\bf Funding}: Alexander von Humboldt (AvH) Stiftung.

\noindent
{\bf Authors' contributions}: Alessandro Berti led the systematic literature review, meticulously summarizing the related work and constructing visual depictions of the various categories. Wil van der Aalst and Marco Montali made valuable contributions by providing insightful suggestions for the formation of the search query used in the literature review, the classification of event logs and process models, and offering critical input to enhance the overall quality of the paper.

\noindent
{\bf Acknowledgements}: We thank the Alexander von Humboldt (AvH) Stiftung for supporting our research.

\clearpage\newpage

\bibliography{new_literature_review}

\end{document}